\documentclass[sigconf]{acmart}

\usepackage{booktabs} % For formal tables
\usepackage[flushleft]{threeparttable}

\setcopyright{rightsretained}
\acmDOI{10.475/123_4}

\acmISBN{123-4567-24-567/08/06}

\acmConference[WOODSTOCK'97]{ACM Woodstock conference}{July 1997}{El
  Paso, Texas USA} 
\acmYear{1997}
\copyrightyear{2016}

\acmPrice{15.00}

\begin{document}
\title{A Deep Multimodal Approach for Cold-start \\Music Recommendation}
%\titlenote{Produces the permission block, and
%  copyright information}
%\subtitle{Extended Abstract}
%\subtitlenote{The full version of the author's guide is available as
%  \texttt{acmart.pdf} document}

\copyrightyear{2017} 
\acmYear{2017} 
\setcopyright{acmlicensed}
\acmConference{DLRS 2017}{August 27, 2017}{Como, Italy}\acmPrice{15.00}\acmDOI{10.1145/3125486.3125492}
\acmISBN{978-1-4503-5353-3/17/08}

\author{Sergio Oramas}
%\authornote{}
\affiliation{%
  \institution{Music Technology Group \\ Universitat Pompeu Fabra}
}
\email{sergio.oramas@upf.edu}

\author{Oriol Nieto}
%\authornote{}
\affiliation{%
  \institution{Pandora Media, Inc.\\Oakland, CA}
}
\email{onieto@pandora.com}

\author{Mohamed Sordo}
%\authornote{}
\affiliation{%
  \institution{Pandora Media, Inc.\\Oakland, CA}
}
\email{msordo@pandora.com}

\author{Xavier Serra}
%\authornote{}
\affiliation{%
  \institution{Music Technology Group \\ Universitat Pompeu Fabra}
}
\email{xavier.serra@upf.edu}

\renewcommand{\shortauthors}{Oramas et al.}

\begin{abstract}
An increasing amount of digital music is being published daily. Music streaming services often ingest all available music, but this poses a challenge: how to recommend new artists for which prior knowledge is scarce? In this work we aim to address this so-called cold-start problem by combining text and audio information with user feedback data using deep network architectures. Our method is divided into three steps. First, artist embeddings are learned from biographies by combining semantics, text features, and aggregated usage data. Second, track embeddings are learned from the audio signal and available feedback data. Finally, artist and track embeddings are combined in a multimodal network. Results suggest that both splitting the recommendation problem between feature levels (i.e., artist metadata and audio track), and merging feature embeddings in a multimodal approach improve the accuracy of the recommendations.
\end{abstract}




\maketitle

\section{Introduction}\label{section:intro}

It is common for online music streaming services nowadays to offer ever-growing catalogs with dozens of millions of music tracks.
Since manually managing these large libraries is not feasible due to size constraints, automatic exploration and exploitation of large-scale music collections has been an active area of research in recent years \cite{schedl2014music}.
While several existing algorithmic techniques are able to produce successful recommendations for popular content \cite{Koren2009}, the exploration of new or \emph{undiscovered} artists (i.e., the long tail \cite{Celma2010}) remains a major challenge that we aim to address.

Recommender systems can be broadly classified into collaborative filtering (CF), content-based, and hybrid methods. CF methods \cite{Koren2009} use the item-user feedback matrix and predictions are based on the similarity of user or item profiles. Matrix factorization techniques are currently CF state-of-the-art \cite{Koren2009}. 
CF methods suffer from the cold-start problem, as new items do not have feedback information \cite{Saveski2014}. Content-based methods \cite{Mooney1999} rely only on item features, and recommendations are based on similarity between such features. Finally, hybrid methods \cite{Burke2002} try to combine both item content and item-user feedback.
%TODO: Background on knowledge-based music recommendation (TODO: cite Netflix, cite w2v).

%TODO: Background on content-based music recommendation (TODO: cite Van den Oord).

Social tags have been extensively used as a source of artist content features to recommend music \cite{Knees2013}. However, these tags are usually collectively annotated, which often introduce an artist popularity bias \cite{Turnbull2008}.
%Este problema también pasa con las biografías, también son colaborative effort
%Artist biographies and press releases are, on the other hand, a more reliable source of context information. They may be produced by artists themselves, or collectively annotated. 
Artist biographies and press releases, on the other hand, do not necessarily require a collaborative effort, as they may be produced by artists themselves. 
However, they have seldom been exploited for music recommendation \cite{Oramas2015}.
Part of this work focuses on learning artist features from these biographies.
Furthermore, we also make use of audio signals, since these are generally always available and have shown to be helpful when recommending music in the long tail \cite{Oord2013}.
% relevant music information can also be extracted directly from the audio signal, which can help to recommend music (TODO: cite van den oord).

According to \cite{gulccehre2016knowledge}, composing simpler tasks is more likely to yield effective local minima for neural networks. In addition, as stated in \cite{larochelle2009exploring}, directly training all the layers of a deep network together make it difficult to exploit all the extra modeling power of a deeper architecture. 
Therefore, we decided to separate the problem of music recommendation into artist and song levels.
Artist feature embeddings are learned from artist metadata in an artist recommendation scenario.
Track feature embeddings are learned from audio signals in a song recommendation scenario.
In both cases, a hybrid recommendation approach is used based on learning attribute-to-feature mappings \cite{GantnerDFRS10}.
This method addresses the lack of feedback for uncommon items in two steps: (1) factorizing the collaborative matrix, and (2) learning a mapping between item content features and item latent factors \cite{Oord2013,Bansal2016}.
Lastly, both feature embeddings are combined in a multimodal network to predict song recommendations of cold-start artists.
We show how dividing the problem into artists and songs, and combining text and audio in a multimodal approach yields improved recommendations.

For the sake of reproducibility, source code and data splits used in the experiments have been released\footnote{https://github.com/sergiooramas/tartarus}.
Our main contributions in this work are summarized as follows:
\begin{itemize}
\item{Method to enrich artist biographies with semantic information leveraging an external Knowledge Base.}
\item{Dividing the problem of music recommendation into artist and song levels to obtain better performance.}
\item{A multimodal deep learning pipeline for music recommendation that combines audio with text and yields improved results.}
\item{The release of an extended version of the Million Song Dataset with artist biographies and tags.}
\end{itemize}

\section{Recommendation Approach}\label{section:approach}

% In this work, a method to recommend items that are new to a recommender system is proposed. 
% More specifically, we aim to address the cold-start problem: where little to none collaborative information is available when producing recommendations.

% This scenario is often referred to as the cold-start problem.
% More specifically, we want to recommend songs whose artist was not previously in the catalog, which means that the system does not have any collaborative information about other songs by the same artist. This problem can be considered as an extreme case of the cold-start scenario.

% To produce music recommendations in the long tail, we propose the following framework.
% Given the set of artist features $A_{s}$ of a song $s$ with $A_{s}={a_{1},a_{2},...,a_{n}}$, and the set of track features $T_{s}$ of $s$ with $T_{s}={t_{1},t_{2},...,t_{m}}$, the complete feature set of $s$ is defined as the aggregation of its artist and track features $F_{s} = A_{s} \cup T_{s}$.
To produce cold-start music recommendations, we propose the following framework.
Given the set of artist features $A_{s}$ of a song $s$, and the set of track features $T_{s}$ of $s$, the complete feature set of $s$ is defined as the aggregation of its artist and track features $F_{s} = A_{s} \cup T_{s}$.
% This aggregated set of features is typically used as input in content-based and hybrid music recommender systems (TODO: cite).

\begin{figure}[!htp]
\centerline{
\includegraphics[width=0.95\columnwidth]{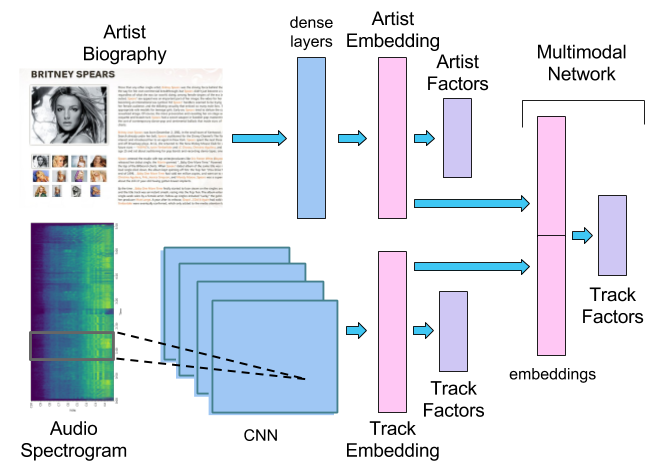}}
\caption{Model architecture.}
\label{fig:approach}
\end{figure}

Given the heterogeneity of these two feature sets (audio and text), a learning process involving them may under-explore one of the modalities, as the stronger modality may dominate quickly. 
To ensure that the variability of the input data is fully represented, we divide the problem into three phases (see Figure~\ref{fig:approach}). First, we aggregate the collaborative information of all songs of the same artist, and learn an artist feature embedding $A'_{s}$ from $A_{s}$ in an artist recommendation scenario. Second, we learn a track feature embedding $T'_{s}$ from $T_{s}$ in a pure audio-based recommendation scenario. Third, we combine both feature embeddings $A'_{s}$ and $T'_{s}$ in a multimodal network and compute song recommendations. 

Since songs from the same artist share the same set $A_{s}$, if different songs from the same artist appear in multiple sets (e.g., train and test), a problem of overfitting may arise \cite{Flexer2007ACL}.
To approach this issue, we use non-overlapping artists across the train, validation, and test sets.

%Since songs from the same artist share the same set $A_{s}$, two problems may arise: (1) the learning process involved with the recommender system might not be properly optimized, and (2) if different songs from the same artist appear in multiple sets (e.g., train and test), a problem of overfitting may arise \cite{Flexer2007ACL}.

%To address (1), we divide the problem into three phases (see Figure~\ref{fig:approach}). First, we aggregate the collaborative information of all songs of the same artist, and learn an artist feature embedding $A'_{s}$ from $A_{s}$ in an artist recommendation scenario. Second, we learn a track feature embedding $T'_{s}$ from $T_{s}$ in a pure audio-based recommendation scenario. Third, we combine both feature embeddings $A'_{s}$ and $T'_{s}$ in a multimodal network and compute song recommendations. To approach (2), we use non-overlapping artists across the train, validation and test sets.

Let $M$ be the matrix of implicit feedback, where $m_{us}$ is the number of play counts for user $u$ on song $s$. 
$M$ is split into $M_{train}$, $M_{val}$ and $M_{test}$, for train, validation and test, respectively, where no artist is shared across sets. 
Factorizing $M_{train}$ using weighted matrix factorization (WMF) \cite{Hu2008} yields $I_{k}$ and $U_{k}$, the $k$ dimensional sets of song and user latent factors, respectively.
We set $k=200$, and apply the alternating least squares (ALS) optimization method.% with the default parameters. 
%, and more explicitly their alternating least squares (ALS) optimization method \cite{}. 

To learn the artist embeddings, we obtain the matrix of artist implicit feedback $R$ from $M$, being $R_{ua} = \sum_{s}m_{us}$ for all songs $s$ from the same artist $a$. This matrix is split into train, validation, and test sets following the same partition of artists made for $M$, and thus keeping the mutual exclusion restriction. 
Latent factors of artists and users are later obtained via WMF. 
Lastly, a deep neural network is trained on the prediction of artist latent factors from artist content features $A$.
On the other hand, the song latent factors are predicted with a deep convolutional network, using $I_{k}$ as training data and the track features $T$ as input (similar to \cite{Oord2013}).
% Given $I_{k}$, a deep neural network is trained on the prediction of song latent factors $I_{k}$ from audio track features $T$.

Once the artist and track models are trained and optimized, we gather the activations from the penultimate layer of each network for all the sets. These activations constitute what we call the artist and track feature embeddings $A'_{s}$ and $T'_{s}$, which are in turn used as input to a third network. 
This final multimodal network is trained on the prediction of song latent factors $I_{k}$ from $S'_{s} = A'_{s} \cup T'_{s}$.
Finally, the list of item recommendations for user $u$ is obtained by ranking the results of computing the dot product between the user latent factor $f_{u} \in U_{k}$ and the set of item factors. %the set of predicted item latent factors in the test set. %This ranking is evaluated in comparison with the ranked list of items for user $u$ present in the test matrix $M_{test}$.

The different architectures used in each one of the three neural networks involved in the approach are described in Sections \ref{section:text} ,\ref{section:audio}, and \ref{section:multimodal}, respectively.
Nevertheless, all networks have a final fully connected layer of 200 units\footnote{to match the dimensions of the factors to be predicted.} with linear activation and l2-normalization. 
In addition, mini batches of 32 items are randomly sampled from the training data to compute the gradient in all networks, and Adam \cite{KingmaB14} is the optimizer used to train the models, with the default suggested learning parameters. 
Given that the output of the architectures are l2-normalized, we use cosine proximity as the loss function, as in \cite{Chollet2016}.

\section{Artist Text Embeddings}\label{section:text}

% Following the approach defined in Section \ref{}, a matrix of artist and songs is obtained by aggregating the play counts of songs of the same artist present in the MSD. Then, a deep network is trained to learn the artist latent factors from the biography texts. 
In this section we describe two different, competing approaches to exploit artist texts in a deep learning process. %Then, these two approaches are evaluated in an artist recommendation scenario. Finally, artist embeddings are obtained.

\subsection{Semantic enrichment}\label{section:sem}

We propose a method for enriching artist biographies by associating text fragments with relevant entities defined in online semantic datasets, and then gathering relevant information about these entities from a Knowledge Base (KB). For this purpose, we adopted Babelfy, a state-of-the-art tool for Entity Linking (EL) \cite{Moroetal2014}, a task to associate, for a given textual fragment candidate, the most suitable entry in a reference KB. Babelfy maps words from a given text to entities in the BabelNet Knowledge Base \cite{Navigli2012}, which has a direct mapping to DBpeadia\footnote{\texttt{http://dbpedia.org}}, a structured version of Wikipedia\footnote{\texttt{http://wikipedia.org}}. We use semantic information about the identified entities coming from DBpedia to enrich the biographies. DBpedia resources are generally classified using the DBpedia Ontology, which is a shallow, cross-domain ontology based on the most common infoboxes of Wikipedia. 

EL systems are useful for music recommendation \cite{Oramas2015}. However, they are not optimized for the music domain, and are prone to errors \cite{Oramas2016a}. The application of a filtering process over the set of identified entities based on their classification within the DBpedia Ontology, has demonstrated its utility to improve music retrieval tasks, such as artist similarity \cite{Oramas2015a}. Therefore, we only keep entities of classes related to the music domain such as MusicalArtist, Band, MusicGenre, MusicalWork, RecordLabel, Instrument, Engineer, and Place. Then, we query DBpedia to get all the available information about the filtered entities. From the information gathered, we keep some specific properties for every kind of entity, such as homeTown, instrument, genre or associatedBand for MusicalArtists, writer, producer or recordedIn for MusicalWorks, stylisticOrigin or instrument for MusicGenres, and so on\footnote{The complete list of classes and properties is available at http://mtg.upf.edu/download/datasets/msd-a}. In addition, we also kept all the Wikipedia categories associated to each entity. In Wikipedia, categories are used to organize resources, and they help users to group articles of the same subject.

To build the enriched biographies we proceed as follows: First, Babelfy is applied over the biography texts. Second, information is gathered from DBpedia for the entities of the selected classes. Finally, the collected data are added at the end of the biography text separated by spaces. A vector space model (VSM) is then applied to the set of enriched biographies, and tf-idf weighting \cite{Zobel1998} is used. We limited the vocabulary size to 10,000 terms for the VSM, as this number provides a good trade-off between performance and number of parameters required for training. Note that either words, entities, dates, or categories may be part of this vocabulary. From this data representation, a feedforward network with two dense layers of 2048 neurons each is trained to predict the artist latent factors.

\subsection{Word embeddings}\label{section:w2v}

Much of the work with deep learning in Natural Language Processing has involved the learning of word vector representations \cite{Bengio2003,Mikolov2013}, and their further composition \cite{Collobert2011}. Word embeddings aim to represent words as low-dimensional dense vectors. They have demonstrated to greatly benefit NLP tasks, such as word similarity, sentiment analysis, or parsing \cite{Nguyen2016}. 

The use of convolutional neural networks (CNN) over pre-trained word vectors has become state-of-the-art in sentence classification \cite{Kim2014}. 
We re-adapt the architecture proposed in \cite{Kim2014} for sentence classification to learn artist latent factors from artist biographies. This consists in an embedding layer, followed by a one dimensional convolutional layer with multiple filter widths, a max-over-time pooling layer, a dense hidden layer and the output layer. We employ the same architecture and parameters, changing only the output layer and the loss function. We initialize the input embedding layer of the network with word2vec word embeddings pre-trained on the Google News dataset, and also with word embeddings trained in our own corpus of biographies.

%The architecture consists in an embedding layer, followed by a one dimensional Convolutional layer with multiple filter widths (2, 3 and 4 in our case). A convolution operation involves a filter, which is applied to a window of h words to produce a new feature. This filter is applied to each possible window of words in a sentence to produce a feature map. Then a max-over-time pooling layer reduce every feature map to a unique feature. The idea is to capture the most important feature with the highest value for each feature map. Finally a fully connected layer precede the output layer, with as many neurons as the latent factor dimension. Dropout is used after every layer to prevent overfitting.

\section{Track Audio Embeddings}\label{section:audio}

It is common in the field of music informatics to make use of CNNs to learn higher-level features from spectrograms.
These representations are typically contained in $\mathbb{R}^{\mathcal{F} \times N}$ matrices with $\mathcal{F}$ frequency bins and $N$ time frames.
In this work we compute 96 frequency bin, log-compressed constant-Q transforms (CQT) \cite{Schorkhuber2010} for all the tracks in our dataset using \texttt{librosa} \cite{Mcfee2015} with the following parameters: audio sampling rate at 22050 Hz, hop length of 1024 samples, Hann analysis window, and 12 bins per octave.
Following a similar approach to \cite{Oord2013}, we address the variability of the length $N$ across songs by sampling one 15-seconds long \emph{patch} from each track, resulting in the fixed-size input to the CNN.
%The labeled data associated to each of these patches is a 200 dimensional vector representing the latent factors obtained by factorizing the user-track matrix using weighted matrix factorization (TODO: cite Hu et al. 2008).
%This matrix contains the implicit feedback represented as play counts for each of the users across all tracks in the collection, and factorizing it has proven to be an effective technique for recommending items (TODO: cite van den Oord, Netflix).

The deep model trained with these data is defined as follows: the CQT patches are fed to four convolutional layers with rectified linear units (ReLU) as activations.
The four convolutions have the following number of filters, from first to last: 256, 512, 1024, and 1024.
The convolutions are only applied to the time axis, leaving the frequencies fixed since the absolute and relative bin placement is important when aiming to capture particular sounds (as opposed to the irrelevance of \emph{where} in time a certain sonic event occurs).
Maxpooling of 4 units across the time axis is applied after each of the first three ReLUs, and 50\% dropout is applied to all layers.
The flattened output of the last layer has 4096 units, which becomes the vector embedding to later use in the multimodal approach described next.
%The final fully connected layer has 200 units (to match the dimensions of the factors aiming to be predicted) with linear activation and l2-normalization.

%Mini batches of 64 patches are randomly sampled from the training data to compute the gradient, and Adam (TODO: cite) is the optimizer used to train the model, with the default suggested learning parameters.
%Given that the output of the architecture is l2-normalized, we use cosine proximity as the loss function.

%260k patches corresponding to the 260k tracks described in \ref{subsec:dataset}, divided into training (80\%), validation (10\%) and test (10\%) sets are used in the training process.
%As opposed to (TODO: cite van den Oord), no artist appears in more than one subset, since it has been shown this could yield overoptimistic results (TODO: cite Arthur Flexer 2007).

% Include architecture figure.

% Discuss hyper-parameters to be explored in section \ref{sec:experim}.

\section{Multimodal Fusion}\label{section:multimodal}

There are several approaches in the literature for multimodal feature learning \cite{ngiam2011multimodal,srivastava2012learning}, and late fusion of multimodal feature vectors \cite{Bechet2015,Slizovskaia2017}. %In this work, feature vectors are concatenated and connected to a dense layer in a fully connected network. 
In this work, audio and text feature vectors are learned separately and then combined via late fusion in a multimodal network (see Figure~\ref{fig:approach}).

Given the different nature of the artist and track embeddings, a normalization step is necessary. 
%Thus, given a set of feature vectors, either batch normalization or $l2$-norm is applied on each of them. 
Normalized feature vectors are then fed to a feed forward neural network (a simple Multi Layer Perceptron, MLP). Two different architectures were explored: (i) each embedding vector is connected to an isolated dense layer of 512 hidden units with ReLU activations after a process of batch normalization \cite{IoffeS15}. Then, both dense layers are connected to the output layer. The rationale behind this is that the isolated dense layers help the network learn non-linearities from each modality separately. (ii) each embedding vector is $l2$-normed and then concatenated into a single feature vector which is directly connected to the output layer, resulting in a linear model.
%(i) single hidden layer of 512 ReLU units and (ii) input directly connected to output with no hidden layers.
%First, each embedding vector is connected to an isolated dense layer of 512 hidden units with ReLU activations. Then, both dense layers are connected to the output layer. Second, both embedding vectors are concatenated into a single feature vector which is directly connected to the output layer, becoming into a linear regression model.
%, obtained the best results in a network where the input layer is directly connected to the output layer (no hidden layer involved). 
%In our configuration, each embedding vector is connected to an isolated dense layer of 512 hidden units with ReLU activations after a process of batch normalization \cite{IoffeS15}. Then, both dense layers are connected to the output layer. The rationale behind this is that the isolated dense layers help the network learn non-linearities from each modality separately. 
Regularization is obtained by applying dropout with an empirically selected factor of 70\% after the input layer for both architectures.

%citar \cite{ngiam2011multimodal} \cite{srivastava2012learning}

\section{Experiments}\label{sec:experim}

\subsection{Dataset}\label{subsec:dataset}

The Million Song Dataset (MSD) \cite{McFee2012} is a collection of metadata and precomputed audio features for 1 million songs. 
Along with this dataset, the Echo Nest Taste Profile Subset \cite{Bertin-Mahieux2011} provides play counts of 1 million users on more than 380,000 songs from the MSD.
Starting from this subset, we gather biographies and social tags from Last.fm for all the artists that have at least one song in the dataset.
When there are several artists with the same name, they are stored in the same page of Last.fm, which makes the biography and social tags ambiguous. 
We automatically removed all ambiguous artists by applying text processing on the biographies. 
The song features provided with the MSD are not generally suitable for deep learning, so we instead use audio previews between 7 and 30 seconds retrieved from \texttt{7digital.com}.
After removing ambiguous artists and missing tracks, the final dataset consists of 328,821 tracks from 24,043 artists. 
Each track has at least 15 seconds of audio, each biography is at least 50 characters long, and each artist has at least 1 tag associated with it. All artist metadata, implicit feedback matrices, and splits are released as a new dataset called the MSD-A\footnote{\url{https://doi.org/10.5281/zenodo.831348}}.

\subsection{Artist Recommendation}
\label{section:artist-rec}

To investigate to what extent the different feature sets, data models and architectures influence the quality of the deep artist features, we evaluate the different approaches in an artist recommendation scenario. Given the matrix of implicit feedback $R$, and the set of artist and user factors obtained through matrix factorization (see Section \ref{section:approach}), we predict the artist factors for the test set, and use them to compute a ranked list of recommended artists for every user. We use mean average precision (MAP) with a cut-off at 500 recommendations per user. 

We compare four different approaches using the biography texts as input. (1) a pure text-based approach using a VSM and a feedforward network \textsc{a-TEXT}. (2) similar to (1) but with a semantically enriched version of the texts \textsc{a-sem} (cf. Section~\ref{section:sem}). (3) An approach based on word embeddings initialized with vectors trained on Google News and a CNN \textsc{a-w2v-goo} (cf. Section~\ref{section:w2v}). (4) Similar to (3) but inizializing the embeddings with word vectors previously trained on the corpus of biographies \textsc{a-w2v}. To properly frame the results, we compute two baselines and one competitor approach. The \textsc{tags} baseline approach uses artist social tags as input features, and \textsc{text-rf} uses biography texts as input, but Random Forest Regression for the learning instead of a deep neural network. The former baseline is added to compare the potential of biography texts with respect to curated metadata, whilst the latter was added to study to which extent the deep network improves the results over other learning methods typically used in natural language processing. There are few recommendation approaches able to deal with an extreme cold-start scenario like ours. Therefore, we select ItemAttributeKnn \cite{GantnerDFRS10} as the competitor approach (\textsc{tags-itemKnn}), using artist social tags as attribute data and computed using the MyMediaLite library\footnote{\texttt{http://www.mymedialite.net/}}. %The input in all the approches is encoded either using a vector space model (VSM) or word embeddings (w2v). %For the VSM input we used a feedforward (FF) network with two hidden layers of 2048 neurons each, whereas for w2v we used the convolutional architecture (CNN) described in Section~\ref{}. 
We also show the scores achieved when the latent factor vectors are randomized (\textsc{random}), and when they are learned from feedback data using matrix factorization (\textsc{upper-bound}).

\begin{table}
\begin{threeparttable}
\centering
\caption{Artist Recommendation Results}
\label{tbl:artists}
\begin{tabular}{lcccl}
\toprule
\textbf{Aproach} & \textbf{Input}   & \textbf{Data model} & \textbf{Arch} & \textbf{MAP} \\ \midrule
\textsc{a-text}                                   & Bio              & VSM                 & FF            & 0.0161                            \\ 
\textbf{\textsc{a-sem}}                           & \textbf{Sem Bio} & \textbf{VSM}        & \textbf{FF}   & \textbf{0.0201}                   \\
\textsc{a-w2v-goo}                                & Bio              & w2v-pretrain        & CNN           & 0.0119                            \\ 
\textsc{a-w2v}                                    & Bio              & w2v-trained         & CNN           & 0.0145                            \\ \midrule
\textsc{a-tags}                                   & Tags             & VSM                 & FF            & 0.0314                            \\ 
\textsc{tags-itemKnn}                                   & Tags             & -                 & itemKnn            & 0.0161                            \\ 
\textsc{text-rf}                                & Bio              & VSM                 & RF            & 0.0089                            \\ \midrule
\textsc{random}                                 & -                & -                   & -             & 0.0014                            \\
\textsc{upper-bound}                            & -                & -                   & -             & 0.5528                            \\ \bottomrule
\end{tabular}
    \begin{tablenotes}
      \small
      \item Mean average precision (MAP) at 500 for the predictions of artist recommendations in 1M users. VSM refers to Vector Space Model, FF to Feedforward, RF to Random Forest, CNN to Convolutional Neural Network, and itemKnn to itemAttributeKnn approach. Bio refers to biography texts and Sem Bio to semantically enriched texts.
    \end{tablenotes}
   \end{threeparttable}
\end{table}

Results reported in Table~\ref{tbl:artists} show that the semantic enrichment of the biographies \textsc{a-sem} outperforms the pure text approach \textsc{a-text}. As expected, the use of tags improves the results over the use of text. However, the addition of semantic features reduces the gap in performance between the use of tags and unstructured text. Moreover, the difference between \textsc{a-text} and \textsc{text-rf} shows that the use of deep learning with respect to random forest improves the results. We also note that a VSM model with a feedforward network outperforms the use of word embeddings with convolutions. Although, according to the literature, this latter approach has demonstrated its utility for simple tasks like binary classification with short texts, our task puts forward two challenges for this architecture: the greater length of the input texts, and the higher dimensionality of the output. Although we have shown that initializing the embedding layer with word vectors trained on the corpus itself (\textsc{a-w2v}) outperforms the use of Google News pre-trained vectors (\textsc{a-w2v-goo}), further work is necessary to properly optimize a convolutional architecture for this task. Finally, we observe that our approach \textsc{a-tags} outperforms the competitor approach \textsc{tags-itemKnn} using the same item attributes.

%\subsection{Artist Embeddings}
Once the network is trained, we predict the activations of the penultimate layer for the entire dataset of artists. Thus, we obtain a vector embedding of 2048 dimensions, which represents the artist deep features $A'$. From the evaluated approaches, we compute the artist embedding from the \textsc{a-sem} and \textsc{a-tags} approaches.

\subsection{Song Recommendation}
\label{section:song-rec}

\begin{table}[]
\begin{threeparttable}
\centering
\caption{Song Recommendation Results}
\label{tbl:song}
\begin{tabular}{lcccl}
\multicolumn{1}{c}{\textbf{Approach}} & \textbf{Artist Input} & \textbf{Track Input} & \textbf{Arch} & \multicolumn{1}{c}{\textbf{MAP}} \\
\toprule
\textsc{audio}                                 & -                     & audio spec        & CNN           & 0.0015                          \\
\textsc{sem-vsm}                               & Sem Bio               & -                    & FF            & 0.0032                          \\
\textsc{sem-emb}                               & \textsc{a-sem}            & -                    & FF            & 0.0034                          \\
\midrule
\textsc{\textbf{mm-lf-lin}}                        & \textsc{\textbf{a-sem}}     & \textbf{\textsc{audio} emb}  & \textbf{MLP} & \textbf{0.0036}                  \\
\textsc{mm-lf-h1}                        & \textsc{\textbf{a-sem}}     & \textsc{audio} emb  & MLP & 0.0035                  \\
\textsc{mm}                                    & Sem Bio               & audio spec        & CNN            & 0.0014                          \\
\midrule
\textsc{tags-vsm}                              & Tags              & -                    & FF            & 0.0043                          \\
\textsc{tags-emb}                              & \textsc{a-tags}           & -                    & FF            & 0.0049                          \\
\midrule
\textsc{random}                                & rnd emb         & -                    & FF            & 0.0002                          \\
\textsc{upper-bound}                           & -                     & -                    & -             & 0.1649                   \\      
\bottomrule
\end{tabular}
\begin{tablenotes}
      \small
      \item Mean average precision (MAP) at 500 for the predictions of song recommendations in 1M users. \textsc{audio} emb refers to the track embedding of \textsc{audio} approach, \textsc{sem} to artist embedding of \textsc{sem} approach, \textsc{tags} to artist embedding of \textsc{tags} approach, spec to spectrogram, mm to multimodal, lf to late fusion, lin to linear, and h1 to one hidden layer.
    \end{tablenotes}
   \end{threeparttable}
\end{table}

In this experiment, audio embeddings are obtained after training the convolutional network (see Section~\ref{section:audio}) with 260k patches of 15 seconds, corresponding to the 80\% of the tracks described in Section~\ref{subsec:dataset}. Patches are divided into training (80\%), validation (10\%) and test (10\%) sets. Results reported in Table~\ref{tbl:song} are computed over the remaining 20\% of tracks. % are used in the training process.
%Then multimodal approaches are computed in the same splits.
As opposed to \cite{Oord2013}, no artist appears in more than one subset to avoid overfitting. Finally, multimodal approaches are computed on the same sets.

In our experiments, we want to measure the impact of the artist embeddings in the song recommendation problem, and also the potential of the multimodal approach. We experimented with two approaches, \textsc{sem-emb} and \textsc{tags-emb}, that exploit the feature embeddings learned from the artists features (see Section~\ref{section:artist-rec}), either based on biography texts (\textsc{a-sem}) or artists tags (\textsc{a-tags}). To measure the potential of the artist embeddings, we also computed two approaches based on the original artist features (\textsc{sem-vsm} for semantic text features and \textsc{tags-vsm} for tag features). Results on Table~\ref{tbl:song} show that \textsc{sem-emb} and \textsc{tags-emb} outperform \textsc{sem-vsm} and \textsc{tags-vsm}, suggesting that artist embeddings outperform artist features.

An approach based on the audio spectrograms was computed (\textsc{audio}). From this latter approach, audio embeddings where obtained (\textsc{audio} emb) and combined with \textsc{a-sem} in a multimodal late fusion approach \textsc{mm-lf-lin} (without hidden layers and $l2$-norm) and \textsc{mm-lf-h1} (with one hidden layer after each feature vector and batch normalization) (cf. Section~\ref{section:multimodal}). We also tried with different combinations of hidden layers and normalization steps in the multimodal network but all of them yielded lower results than the ones reported for \textsc{mm-lf-lin} and \textsc{mm-lf-h1}. We compared this network with a multimodal approach trained directly on the original features (semantically enriched text and audio spectrograms). Results on the combination of artist and track features show that the late fusion of artist and track embeddings \textsc{mm-lf-lin} outperforms the simultaneous training of artist and track features \textsc{mm}. In addition, we observe that we achieve better results when no hidden layer is added to the multimodal network \textsc{mm-lf-lin}. Finally, we observe that the multimodal approach that combines text and audio features with late fusion \textsc{mm-lf-lin} improves the results of pure text \textsc{sem-emb} or pure audio \textsc{audio} approaches. All the differences between the approaches are statistically significant ($p < 0.01$) according to the paired $t$-test.

We also compared the results with an upper-bound approach obtained from the feedback data and an approach trained with random vector embeddings. Although results are in general far from the upper-bound, the comparative analysis of the proposed approaches gives some insights of the behavior of different feature representations and modalities in the cold-start recommendation problem.

%\section{Related Work}

\section{Conclusions}

In this work, a multimodal approach for song recommendation has been presented. The approach is divided into three steps. (1) Artist feature embeddings are learned from text and semantic features in an artist recommendation scenario using a deep network architecture. (2) Track feature embeddings are learned from the audio spectrograms using convolutional neural networks. (3) Embeddings are combined in a multimodal network.

Results show that splitting the problem of song recommendation at artist and song levels improves the quality of recommendations. 
Learning artist feature embeddings separately benefits from the aggregation of the information about the different songs of the same artist, yielding more robust artist features. 
% This approach led to the creation of more powerful artist features. 
Related to this, an approach for the semantic enrichment of artist metadata has been proposed, leading to a significant improvement in the results. 
In addition, we have shown the potential of exploiting artist biographies in music recommendation. 
Moreover, the deep learning architectures of this work have demonstrated their capacity to improve upon other learning models under the music recommendation framework. 
Finally, we have shown how a multimodal approach, based on the late fusion of track and artist feature embeddings that are learned separately, outperforms end-to-end multimodal approaches where the different modalities are learned simultaneously. Moreover, results have shown that our multimodal approach achieves better results than pure text or audio approaches. 
As future work, we plan to do fine-tunning of the combined model after pre-training the network on each modality separately.

\begin{acks}
This work was partially funded by the Spanish Ministry of Economy and Competitiveness under the Maria de Maeztu Units of Excellence Programme (MDM-2015-0502).
The Tesla K40 used for this research was donated by the NVIDIA Corporation.

\end{acks}

\bibliographystyle{ACM-Reference-Format}
\bibliography{sample-bibliography} 

\end{document}